\documentclass[aps,prd,reprint,superscriptaddress,floatfix]{revtex4-1}
\usepackage{color}
\usepackage{natbib}
\usepackage{graphicx}
\usepackage{amsmath}
\usepackage{layouts}
\usepackage{tabularx}
\usepackage{array}

\begin{document}

\title{Search for microwave emission from ultrahigh energy cosmic rays}
\date{Received 25 May 2012; published 26 September 2012}
\doi{10.1103/PhysRevD.86.051104}

\author{J.~Alvarez-Mu\~{n}iz} 
\affiliation{Universidad de Santiago de Compostela, Departamento de F\'{i}sica de Part\'{i}culas, Campus Sur, Universidad de Santiago, E-15782 Santiago de Compostela, Spain}

\author{A.~Berlin}
\affiliation{Kavli Institute for Cosmological Physics, University of Chicago, 5640 South Ellis Avenue, Chicago, IL 60637, USA} 

\author{M.~Bogdan}
\affiliation{Kavli Institute for Cosmological Physics, University of Chicago, 5640 South Ellis Avenue, Chicago, IL 60637, USA}

\author{M.~Boh\'{a}\v{c}ov\'{a}}
\affiliation{Kavli Institute for Cosmological Physics, University of Chicago, 5640 South Ellis Avenue, Chicago, IL 60637, USA} 
\affiliation{Institute of Physics of the Academy of Sciences of the Czech Republic, 182 21 Prague 8, Czech Republic} 

\author{C.~Bonifazi}
\affiliation{Universidade Federal do Rio de Janeiro, Instituto de F\'{i}sica Cidade Universit‡ria, Caixa Postal 68528, 21941-970, Rio de Janeiro, Brazil}

\author{W.~R.~Carvalho~Jr.}
\affiliation{Universidad de Santiago de Compostela, Departamento de F\'{i}sica de Part\'{i}culas, Campus Sur, Universidad de Santiago, E-15782 Santiago de Compostela, Spain}

\author{J.~R.~T.~de~Mello~Neto}
\affiliation{Universidade Federal do Rio de Janeiro, Instituto de F\'{i}sica Cidade Universit‡ria, Caixa Postal 68528, 21941-970, Rio de Janeiro, Brazil}

\author{P.~Facal~San~Luis}
\affiliation{Kavli Institute for Cosmological Physics, University of Chicago, 5640 South Ellis Avenue, Chicago, IL 60637, USA} 

\author{J.~F.~Genat}
\affiliation{Kavli Institute for Cosmological Physics, University of Chicago, 5640 South Ellis Avenue, Chicago, IL 60637, USA} 

\author{N.~Hollon}
\affiliation{Kavli Institute for Cosmological Physics, University of Chicago, 5640 South Ellis Avenue, Chicago, IL 60637, USA} 

\author{E.~Mills} 
\affiliation{Kavli Institute for Cosmological Physics, University of Chicago, 5640 South Ellis Avenue, Chicago, IL 60637, USA} 

\author{M.~Monasor}
\affiliation{Kavli Institute for Cosmological Physics, University of Chicago, 5640 South Ellis Avenue, Chicago, IL 60637, USA} 

\author{P.~Privitera}
\affiliation{Kavli Institute for Cosmological Physics, University of Chicago, 5640 South Ellis Avenue, Chicago, IL 60637, USA} 

\author{L.~C.~Reyes}
\affiliation{Kavli Institute for Cosmological Physics, University of Chicago, 5640 South Ellis Avenue, Chicago, IL 60637, USA}
\affiliation{Department of Physics, California Polytechnic State University, San Luis Obispo, CA 93401, USA}

\author{B.~Rouille~d'Orfeuil}
\affiliation{Kavli Institute for Cosmological Physics, University of Chicago, 5640 South Ellis Avenue, Chicago, IL 60637, USA}

\author{E.~M.~Santos}
\affiliation{Universidade Federal do Rio de Janeiro, Instituto de F\'{i}sica Cidade Universit‡ria, Caixa Postal 68528, 21941-970, Rio de Janeiro, Brazil}

\author{S.~Wayne}
\affiliation{Kavli Institute for Cosmological Physics, University of Chicago, 5640 South Ellis Avenue, Chicago, IL 60637, USA} 

\author{C.~Williams}
\email[Corresponding Author:~~]{christopherw@uchicago.edu}
\affiliation{Kavli Institute for Cosmological Physics, University of Chicago, 5640 South Ellis Avenue, Chicago, IL 60637, USA}

\author{E.~Zas}
\affiliation{Universidad de Santiago de Compostela, Departamento de F\'{i}sica de Part\'{i}culas, Campus Sur, Universidad de Santiago, E-15782 Santiago de Compostela, Spain}

\author{J.~Zhou}
\affiliation{Kavli Institute for Cosmological Physics, University of Chicago, 5640 South Ellis Avenue, Chicago, IL 60637, USA}

\begin{abstract}
We present a search for microwave emission from air showers induced by ultrahigh energy cosmic rays with the microwave detection of air showers experiment. No events were found, ruling out a wide range of power flux and coherence of the putative emission, including those suggested by recent laboratory measurements.
\end{abstract}

\maketitle
\section{Introduction} 
Evidence for radio emission in the GHz regime from test-beam induced air showers by Gorham \textit{et al.}~\cite{PhysRevD.78.032007} suggests a novel technique to detect extensive air showers (EAS) produced in the atmosphere by ultra-high energy cosmic rays (UHECRs). 

Microwave bremsstrahlung emission~\cite{PhysRevD.78.032007} is expected to be isotropic and follow the EAS longitudinal development, providing an analog to the already successful fluorescence detection technique~\cite{FD,PAOfd}, but with a tenfold increase in detector duty cycle and the elimination of atmospheric systematics. In addition, thanks to the commercialization of the GHz frequency band, off-the-shelf components are readily available, allowing for inexpensive instrumentation of very large areas. A new detector of this kind could allow current experiments like the Pierre Auger Observatory~\cite{PAO} and Telescope Array~\cite{TA}, as well as a future larger scale EAS observatory, to collect a statistically significant sample of UHECR measurements sensitive to composition at the highest energies which is crucial to unveil the origin of UHECRs~\cite{RevModPhys.83.907}. 

Several complementary approaches to microwave detection of EAS are currently being pursued, including the AMBER, EASIER and MIDAS detectors~\cite{GHZAuger} at the Pierre Auger Observatory, and the CROME experiment at KASCADE~\cite{CROME}. Also, new laboratory measurements are being performed~\cite{MAYBE,AMY}.
Prior to its installation at the Pierre Auger Observatory, we have built and operated at the University of Chicago campus a prototype of the MIcrowave Detection of Air Showers (MIDAS) detector~\cite{midasNIMpreprint}, which is sensitive enough to detect EAS microwave signals as estimated from measurements in~\cite{PhysRevD.78.032007}. 
In this Letter, we report results of this first period of data taking and the corresponding limits on the power flux and coherence of microwave emission from EAS. 

\section{MIDAS Detector}
The MIDAS detector consists of a 4.5 m diameter motorized parabolic reflector deployed at the University of Chicago, instrumented with a 53-pixel prime-focus receiver array built from commercially sourced low noise block feeds (LNBFs, C-Band 3.4-4.2 GHz). The corresponding field of view is approximately $20^\circ \times 10^\circ$, with an average system temperature of 65 K. Each pixel's signal is passed through a band-pass filter which rejects most of the anthropogenic noise, fed into an analog power detector and then digitized by a 20 MHz analog-to-digital converter (ADC).  Self-triggering of the MIDAS detector is implemented through field programmable gate arrays (FPGA) in digital electronics boards. For each pixel, a first level trigger (FLT) is issued when a 1 $\mu$s ADC running sum over the pixel's data time stream is over threshold. The pixels' thresholds are dynamically regulated to keep the  FLT rate close to 100 Hz. The second level trigger (SLT) searches for 4-pixel patterns of FLTs with a track-like topology across the camera and overlapping in time within the 10 $\mu$s FLT time gate, as expected of EAS induced signals. When a SLT is found, a 100 $\mu$s ADC time trace is written to disk for each pixel of the detector. SLT rates during data taking varied between 0.01 Hz and 2 kHz depending on noise conditions.  A data taking run lasts six hours, and is automatically restarted. Remote operation and monitoring of the MIDAS detector is implemented.  Details can be found in ~\cite{midasNIMpreprint,Mircea}.

\begin{table*}[t]
\caption{\label{cutTable}Table of cuts used in search program and their effect on selected data sample}
\begin{center}
\begin{tabularx}{0.9\textwidth}{cp{0.77\textwidth}r}
\toprule
\multicolumn{1}{c}{Cut}&\multicolumn{2}{r}{Events Remaining After Cut}\\
\hline 
(1)&	Less than 3 FLT pixels outside the SLT time window 						&	625,012\\
(2)&	All SLT  patterns are time-ordered down-going								&	4,112\\
(3)&	SLT pattern crossing time greater than 400 ns 								&	1,432\\
(4)&	Traces in triggered SLT patterns contain only 1 pulse $>5~\sigma$				&	979\\
(5)&	Pulses $>5~\sigma$ have a shape consistent with power detector's time constant 	&	924\\
(6)&	FLT pixels matching a 5-pixel pattern topology with down-going time order 		&	21\\
(7)&  Visual inspection of candidate events 									&	0\\
\hline
\end{tabularx}
\end{center}
\label{tab:table1}
\end{table*}

\section{Microwave signal simulation}
The sensitivity of the MIDAS detector to EAS has been estimated with an end-to-end Monte Carlo simulation which includes a full description of the antenna characteristics.  Following~\cite{PhysRevD.78.032007}, we parameterize in the simulation the microwave flux of the EAS at the MIDAS detector, $I_{\rm{f}}$, as:
\begin{equation}
\label{eq:microwave_flux}
I_{\rm{f}}=I_{\rm{f,ref}} \cdot \left(\rho/\rho_{0}\right) \cdot \left(d/R\right)^{2} \cdot \left(N/N_{\rm{ref}}\right)^{\alpha}
\end{equation}
where $I_{\rm{f,ref}}$ is the power flux at a distance $d=0.5$~m from a reference shower of $E_{ref} = 3.36\times10^{17}$~eV as given by~\cite{PhysRevD.78.032007}, $R$ is the distance between the detector and the EAS segment, $\rho~(\rho_{0})$ is the atmospheric density at the altitude of the EAS segment (at sea level), and $N$ is the number of shower particles in the EAS segment.  $N_{\rm{ref}}$ is the average number of shower particles at the maximum of the EAS development for a proton primary of energy $E_{ref}$.  $N$ and $N_{\rm{ref}}$ are given by a Gaisser-Hillas~\cite{GH} parameterization of their respective EAS. The parameter $\alpha$ characterizes phenomenologically the coherence scaling relationship for the EAS microwave emission. 

\begin{figure}[h]
\centering
\includegraphics[width=0.45\textwidth]{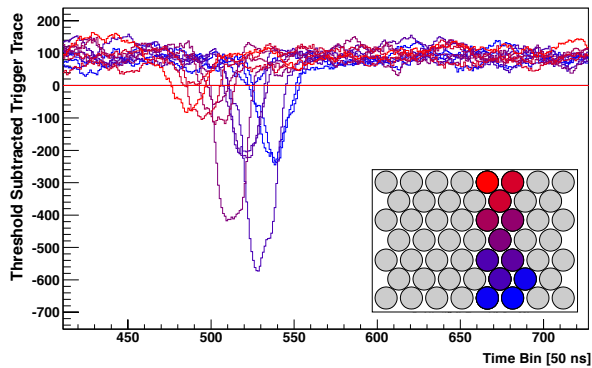}
\caption{Simulated signal from an UHECR shower of energy  $3\times10^{18}\ \rm{eV}$  at 5.5 km distance from the MIDAS detector with $I_{\rm{f,ref}}=1.85\times10^{-15}\ \rm{W/m^2/Hz}$ and quadratic scaling, $\alpha=2$. The histogram shows the FLT running sum calculated over the time trace for the triggered pixels of the camera (shown in the insert), with the horizontal line indicating the FLT threshold level.}
\label{fig:sim}
\end{figure}
The microwave flux received at the MIDAS detector is propagated through the geometry of the reflector and LNBFs according to a microwave optics simulation, and converted to an ADC value. 
The gain pattern of the antenna used in the simulation was cross-checked with measurements of astrophysical GHz sources passing in the field of view. Absolute calibration of the MIDAS sensitivity with solar flux measurements was used for the ADC conversion.   The simulated ADC signal is then processed using software triggering with the same logic implemented in the MIDAS FPGA~\cite{midasNIMpreprint}.

Coherent plasma emission can occur by the superposition of emission wavefronts caused by the spatially coherent distribution of individual emitters.  
In this work, we used a phenomenological parameterization of the plasma coherence which may account for the different conditions present in an EAS and in the laboratory measurements.  We assume a plasma of $N_{\rm{e}}$ electrons to be composed of $M$ subgroups of $\mu_{\rm{e}}$ coherent emitters, $N_{\rm{e}}=M \mu_{\rm{e}}$.  The total emitted power is then given by the incoherent sum of the emission from $M$ subgroups: 
\begin{equation}
I_{\rm{f,tot}}=M \mu_{\rm{e}}^{2} I_{\rm{f,1}}, 
\label{eq:coherent1}
\end{equation}
\noindent where $I_{\rm{f,1}}$ is the power flux from a single electron in the plasma.   Equation \ref{eq:coherent1} can be written as
\begin{equation}
I_{\rm{f,tot}}=N_{\rm{e}}^{\alpha} I_{\rm{f,1}},
\label{eq:coherent2}
\end{equation}
\noindent with the scaling parameter $\alpha=1+\frac{\log \mu_{\rm{e}}}{\log N_{\rm{e}}}$.
The number of coherent emitters, $\mu_{\rm{e}}$, and thus the value of $\alpha$, will depend on the coherent spatial properties of the plasma, including its temperature, density, and electron energy distribution, which may significantly differ between an EAS induced plasma and a plasma generated with test-beams in laboratory measurements. In any case, a power law dependence as in Eq. \ref{eq:microwave_flux} is expected, from Eq. \ref {eq:coherent2} and the proportionality of the number of plasma electrons to the number of shower particles.  Since there is little understanding, both theoretically and experimentally, of the properties of the tenuous air plasma generated by EAS,  we have no definite prediction for $\alpha$, which may take any value from one (incoherent emission) to two (fully coherent emission). When comparing simulations to data, the degeneracy in Eq.~\ref{eq:microwave_flux} between $I_{\rm{f,ref}}$ and the coherence scaling parameter $\alpha$ will be taken into account.  

Laboratory measurements~\cite{PhysRevD.78.032007} suggest a reference power flux  $I_{\rm{f,ref}}^0=1.85\times10^{-15}\ \rm{W/m^2/Hz}$ and full coherence at shower maximum, $\alpha=2$. With these values of emission parameters, MIDAS should detect several tens of highly distinctive events per month. An example of a simulated event is shown in Fig. \ref{fig:sim}.

\section{Data Sample and Event Selection}
The MIDAS detector took data during several months in 2011, under various conditions of anthropogenic noise and with several periods dedicated to calibration measurements.  To ensure quiet and stable data taking conditions, we limited the search for EAS candidates to data runs  with an average SLT rate less than 0.7 Hz. This requirement eliminates data taking periods with particularly noisy conditions. The final data sample contains $1.1 \times10^6$  SLT events collected during 61 days of live time.

We expect only $\approx 1600$ SLT events from accidental coincidences of background fluctuations, indicating the data sample is likely dominated by anthropogenic noise.  Simple event selection criteria, based on timing and trigger information expected from EAS geometries and laboratory measurements of detector electronics, were found to eliminate most of the noise events. Table \ref{tab:table1} summarizes the selection criteria, and the corresponding number of selected events.  

With cut  \textbf{(1)}, we require that no more than two FLT pixels in the event not participating in any SLT pattern are triggered outside a time window defined by the smallest and the largest FLT time amongst the pixels forming SLTs. This cut rejects a class of background events characterized by a large number of camera pixels having a FLT, but distributed in time outside the SLT window and having topologies that do not form a SLT pattern.

A large background rejection is obtained by asking that the time ordering of the FLT pixels is consistent with a down-going track topology for triggered SLTs, as expected for EAS (cut \textbf{(2)}).
Also, we require the FLT time difference between the latest and the earliest pixel of the SLT pattern to be larger than 400 ns (cut \textbf{(3)}), in order to reject a class of background events with coincident pulses within the typical response time of the power detector.  Cut \textbf{(3)} will also eliminate cosmic ray events which produce fast signals, like those with geometry pointing towards the telescope, which would be in any case very difficult to distinguish from the overwhelming background of fast anthropogenic transients.  In addition,  each pixel participating in the SLT must  have recorded only one pulse $>5~\sigma$ (cut \textbf{(4)}), to reject events with multiple pulses in the time trace, typical of noise induced by airplane altimeters and communications. With cut \textbf{(5)}, we ensure  that the detected pulse in each SLT pixel time trace is consistent with an RF pulse passing through the power detector - a pulse with a fall time shorter than the $\rm{33\ ns}$ time constant of the power detector is likely to be a random fluctuation or anthropogenic noise.
Finally, we search the events for a 5-pixel track topology with  down-going time ordering (cut \textbf{(6)}). We expect 0.013 events from accidental coincidences with such 5-pixel topology in 61 days of live time, which would make any candidate event highly significant. Only 21 events survive this cut. 

\begin{figure}[t]
\centering
\includegraphics[width=0.45\textwidth]{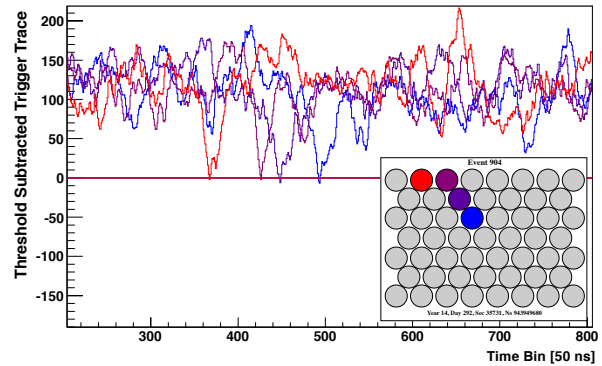}
\caption{One of the candidate 4-pixel events found in the data exhibiting topology which is consistent with expectations from an EAS.}
\label{fig:candidate}
\end{figure}

A visual inspection of the selected events was then performed  (cut \textbf{(7)}). All were identified as clear anthropogenic noise events, illuminating large portions of the camera with multiple coincident pulses that fall below the 5~$\sigma$ threshold in cut \textbf{(4)}. Background events with similar topology and a stronger signals are seen frequently in the data and are indeed rejected by our selection criteria.  The most likely explanation is that the residual background events after cut \textbf{(6)} are produced by the same source of microwave radiation, but with a different geometry relative to the detector.  Notice that almost all of the 4-pixel events remaining after  cut \textbf{(5)} are background, as evident after a visual scan of the candidates.  One example of a 4-pixel event compatible with an EAS signal is shown in Fig.~\ref{fig:candidate}. However, the characteristics of these events, which have signals close to the trigger threshold, are also compatible with a tail of the overwhelming background noise. No strong conclusion can be drawn on the origin of these kinds of events until a validation by a coincident detection at the Pierre Auger Observatory is performed.  Thus, we use the stronger 5-pixel selection, which yielded a null result, to establish a limit on microwave emission from EAS.   Notice that the large field of view of the MIDAS telescope and our search criteria, based only on topology and timing of the signals and with no requirement on the position of the maximum development of the shower, makes the result of our search insensitive to the composition of UHECRs.

\section{Results}\label{results}
In order to establish a limit on the microwave emission from UHECR induced air showers we simulated a large sample of Monte Carlo events spanning a range of coherence scaling parameter $\alpha$ values between one and two, and of reference power flux $I_{\rm{f,ref}}$ values between $2.3\times10^{-16}\ \rm{W/m^2/Hz}$ to $4.6\times10^{-15}\ \rm{W/m^2/Hz}$ (cf.~Eq.~\ref{eq:microwave_flux}) for the MIDAS configuration operating at the University of Chicago campus.  
For each pair of  $\{\alpha,I_{\rm{f,ref}}\}$,  events are simulated in discrete bins of primary energies between $\log _{10} E=17.65$ and $\log _{10} E=20.05$ in logarithmic steps of 0.1, sampled from an isotropic distribution.  For each energy bin, the simulation is run until $4000$ SLTs or $4\times10^{6}$ simulated EAS are reached, whichever occurs first.  The result is then weighted with the UHECR energy spectrum of \cite{Abraham2010239} to produce a realistic energy spectrum of simulated SLT events. 

\begin{figure}[t]
\centering{
\includegraphics[width=0.45\textwidth]{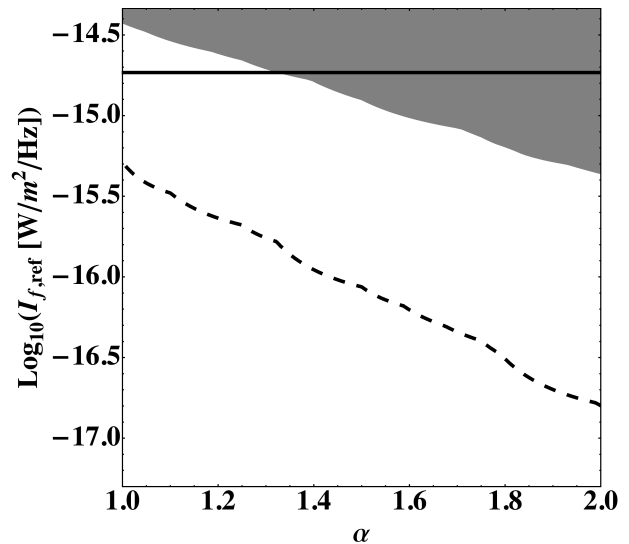}}
\caption{Exclusion limits on the microwave emission from UHECRs, obtained with 61 days of live time measurements with the MIDAS detector.  The power flux  $I_{\rm{f,ref}}$ corresponds to a reference shower of   $3.36\times10^{17}$eV,  and the $\alpha$ parameter characterizes the possible coherence of the emission. The shaded area is excluded with greater than 95\% confidence.  The horizontal line indicates the reference power flux suggested by laboratory measurements~\cite{PhysRevD.78.032007}. The projected 95\% CL sensitivity after collection of one year of coincident operation data of the MIDAS detector at the Pierre Auger Observatory is  represented by the dashed line.}
\label{fig:exclusion}
\end{figure}

The selection criteria \textbf{(1)} to \textbf{(6)} were applied to the simulated samples, and corresponding number of expected events were derived. For example,  for a fully coherent emission $( \alpha=2 )$  and a reference power flux  $I_{\rm{f,ref}}=I_{\rm{f,ref}}^0$, we expect, in our 61 days of livetime, greater than 15 EAS candidate events above $2\times10^{18}$~eV passing all our search criteria.  This is excluded at a greater than $5~\sigma$ level by our null detection result. To cover the full space of $\{\alpha,I_{\rm{f,ref}}\}$, we used a linear surface interpolation of the grid of simulated samples.  The shaded gray region in Fig.~\ref{fig:exclusion} is excluded with greater than 95\% confidence.  We exclude the reference power flux  $I_{\rm{f,ref}}^0$ as measured by~\cite{PhysRevD.78.032007}, indicated by the solid horizontal line in  Fig.~\ref{fig:exclusion}, over a large range of partial coherence  emission hypotheses. An incoherent emission for that reference power flux  is not yet excluded by our measurement. 

\section{Conclusions}
We reported the results of a search for microwave emission from UHECRs  with the MIDAS experiment.  No candidate events were found in the 61 days of live time of the measurement, resulting in the strongest limits to date on the power flux and coherence of EAS microwave emission.  In particular,  we rule out fully quadratic scaling at the power flux value measured by Gorham \textit{et al.}~\cite{PhysRevD.78.032007}. This result has implications on the prospects for the microwave detection of UHECRs, indicating a higher detection energy threshold (or the need for a larger antenna) than the one so far considered. Notice that our limits apply to any microwave emission mechanism producing isotropic radiation that can be detected from the side of EAS at large distances, as in the fluorescence detection technique. The time response of the MIDAS electronics and the event search strategy is not optimized for detection of very fast signals that may come from a beamed emission along the shower axis, as potentially observed in preliminary reports of EASIER~\cite{GHZAuger} and CROME~\cite{CROME}.   

Measurements will continue, with  the MIDAS detector being deployed at the Pierre Auger Observatory in Malarg\"{u}e, Argentina.  Operating the MIDAS experiment in coincidence with the Auger Observatory will provide a much greater sensitivity to EAS events over the search reported here, eliminating the need for severe topological and timing cuts on candidate events.  In addition, the rural environment of the site is significantly more radio quiet than the urban environment of the University of Chicago campus, increasing overall detector livetime. The projected 95\% CL sensitivity after one year of data is collected at the Pierre Auger Observatory is shown as a dashed line in Fig.~\ref{fig:exclusion}, indicating that exclusion of incoherent emission over a large range of reference fluxes will be achieved. 

\section{Acknowledgements}
This work was supported in part by the Kavli Institute for Cosmological Physics at the University of Chicago through grants NSF PHY-0114422 and NSF PHY-0551142 and an endowment from the Kavli Foundation and its founder Fred Kavli, Conselho Nacional de Desenvolvimento Cient'fico e Tecnol—gico (CNPq), Xunta de Galicia (INCITE09 206 336 PR), Ministerio de Educaci\'on, Cultura y Deporte (FPA 2010Ð18410), ASPERA (PRI-PIMASP-2011-1154) and Consolider CPAN Ð Ingenio2010, Spain.  The simulations used in this work have been performed on the Joint Fermilab - KICP Supercomputing Cluster, supported by grants from Fermilab, Kavli Institute for Cosmological Physics, and the University of Chicago.

We would like to thank Peter Gorham, Antoine Letessier-Selvon, and Ralph Engel for insightful discussions on the topic of microwave emission from EAS.

\end{document}